\documentstyle[12pt,preprint]{aastex}  

\def\kms{{\rm\,km\,s^{-1}}}
\def\kmskpc{{\rm\,km\,s^{-1}{kpc}^{-1}}}

\def\deg{{^\circ}}

\def\kpc{{\rm\,kpc}}

\def\mathnew{\mathsurround=0pt}   
\def\simov#1#2{\lower .5pt\vbox{\baselineskip0pt  
    \lineskip-.5pt\ialign{$\mathnew#1\hfil##\hfil$\crcr#2\crcr\sim\crcr}}}

\def\'#1{\ifx#1i{\accent"13\i}\else{\accent"13#1}\fi}    
  
\def\et{et~al. }     
 
\begin{document}    
\shorttitle{Pitch Angle in Late Type Spiral Galaxies and Orbital Chaos}  
\shortauthors{P\'erez-Villegas et al. 2011}

\title{Pitch Angle Restrictions in Late Type Spiral Galaxies Based on Chaotic and Ordered Orbital Behavior}

\author{A. P\'erez-Villegas$^{1}$, B. Pichardo$^{1}$, E. Moreno$^{1}$, A. Peimbert$^{1}$, H. M. Vel\'azquez$^{2}$}     
 
\affil{$^{1}$Instituto de Astronom\'ia, Universidad Nacional
Aut\'onoma de M\'exico, A.P. 70-264, 04510, M\'exico, D.F.; Universitaria, D.F., M\'exico; \\ $^{2}$Observatorio Astron\'omico Nacional, Universidad Nacional Aut\'onoma de M\'exico, Apdo. postal 877, 22800 Ensenada, M\'exico; \\ barbara@astroscu.unam.mx}

\begin{abstract}
We built models for low bulge mass spiral galaxies (late type as
defined by the Hubble classification) using a 3-D self-gravitating
model for spiral arms, and analyzed the orbital dynamics as a function
of pitch angle, going from 10$\deg$ to 60$\deg$. Testing undirectly
orbital self-consistency, we search for the main periodic orbits and
studied the density response. For pitch angles up to approximately
$\sim 20\deg$, the response supports closely the potential permitting
readily the presence of long lasting spiral structures. The density
response tends to ``avoid'' larger pitch angles in the potential, by
keeping smaller pitch angles in the corresponding response. Spiral
arms with pitch angles larger than $\sim 20\deg$, would not be
long-lasting structures but rather transient. On the other hand, from
an extensive orbital study in phase space, we also find that for late
type galaxies with pitch angles larger than $\sim 50\deg$, chaos
becomes pervasive destroying the ordered phase space surrounding the
main stable periodic and quasi-periodic orbits and even destroying
them. This result is in good agreement with observations of late type
galaxies, where the maximum observed pitch angle is $\sim 50\deg$.
\end{abstract}


\section{Introduction}                                                     
\label{sec:intro}    
It is assumed that the Hubble sequence reveals a strong correlation
between the morphology of galaxies and their formation process. The
study of this relation is a very active field where the merger
hypothesis plays a fundamental role; however, it is very unlikely that
all trends observed in this sequence can be explained solely by this
hypothesis. In particular, the long-term galactic dynamics of
non-interacting galaxies is also determined by their inner
structure. These highly non-linear stellar systems are prone to
exhibit the coexistence of an astonishing dichotomy, an exquisite
order right together with counterintuitive complex regions of chaos,
where non-axisymmetric features such as spiral arms, bars, etc, play a
key part.

Regarding to ordered motion, smooth and weak long-lasting large scale
structures such as spiral arms (in general, low mass and/or low pitch
angle), when modeled, present a relatively simple orbital structure
made of families of quasi-periodic orbits surrounding the main
periodic obits that sculpt spiral arms. The density response is also
smooth and coincides nicely with the imposed potential in the case of
low pitch angle spiral arms, for example. There are indications that
even chaotic orbits could reinforce observed morphological features in
this case (Kaufmann \& Contopoulos 1996; Patsis, Athanassoula \&
Quillen 1997; Harsoula, Kalapotharakos \& Contopoulos 2011). However,
orbital analysis on dynamical models suggests that chaotic motion
plays a significant role (Contopoulos 1983,1995; Contopoulos,
Varvoglis \& Barbanis 1987; Grosb\o l 2003) in spirals.

Indeed, large scale structures are not expected to emerge on systems
built out of pure chaos, this is, as long as chaos does not become
pervasive, large scale structures of discs are practically
unaffected. Recently, there has been an interesting discussion about
the possible chaotic nature of the spiral structure (Patsis 2006;
Voglis, Stavropoulos \& Kalapotharakos 2006; Romero-G\'omez \et 2007;
Voglis, Tsoutsis \& Efthymiopoulos 2006; Contopoulos \& Patsis 2008;
Patsis \et 2009).

Although the best known part of the Hubble classification regarding
pitch angles, categorizes galaxies assuming that late types possess
the most opened spiral arms (largest pitch angles), this is just the
envelope of the classification. Late type spirals actually present a
large scatter in this parameter (Kennicutt 1981; Ma \et 2000), going
from about 10$\deg$ to 50$\deg$. In particular, late type spirals (Sb
to Scd), are better fitted with strong spirals (Contopoulos \& Grosb\o
l 1986; Patsis et al. 1991; Patsis, Grosb\o l \& Hiotelis 1997, and
references therein), meaning they are far from being a slight
perturbation that can be reproduced by the Lin \& Shu (1967) spiral
arm potential with a cosine function, solution of the Tight Winding
Approximation (TWA).

We present a first result of a detailed orbital study on models of
late type spiral galaxies as defined by Hubble (1926). In particular,
this work is devoted to one of the structural parameters of spiral
arms: the pitch angle. In this study some restrictions are imposed
theoretically on their steady or transient nature, and on their
maximum pitch angle.

This letter is organized as follows. In Section \ref{model}, the 3-D
galactic potential used to compute orbits is briefly described. In
Section \ref{results} we present our results with periodic orbit
analysis, density response and phase space studies. Finally, in
Section \ref{conclusions} we present our conclusions.

\section{Methodology and Numerical Implementation}\label{model} 
A common method to study the effect of spiral arms on stellar dynamics
recurs to the use of elegant but simple 2-D bisymmetric local
approximations such as cosine functions (TWA based), in this scheme,
spiral arms are assumed to be smooth self-consistent perturbations to
the axisymmetric potential. Cosine potentials to represent spiral
arms, are in several cases taken beyond its self-consistent validity
regime by imposing large pitch angles and/or large amplitudes for the
spiral arms. However, in this regime, other methods to test undirectly
self-consistency of steady potentials, like the construction of
periodic orbits, are applicable. Details of a non-axisymmetric
potential are not negligible when we are talking about a global model,
about sensitive material as it is the gas, or about sensitive orbital
behavior as chaos.  Fine details of a complicate three dimensional
distribution as spiral arms in galaxies are far beyond of an
approximation as simple as a cosine function. Thus, we use a better
approximation based on a 3-D model density distribution, that allows
more complicate shapes and a more detailed representation of a
potential for spiral arms.

We employ the spiral arms potential formed by oblate spheroids called
{\tt PERLAS} (S{\bf p}iral arms pot{\bf e}ntial fo{\bf r}med by ob{\bf
  la}te {\bf s}pheroids) from Pichardo \et (2003), to represent their
3-D density based nature. This 3-D steady two-armed self-gravitating
potential results to be more realistic in the sense that it considers
the force exerted by the whole spiral arms structure, sculpting much
more complicated shapes for the gravitational potential and
gravitational force than a simple 2-D cosine function. This intrinsic
difference gives rise to significant deviations on orbital dynamics
when compared to a cosine potential. Comparisons of the model with
observations and with other models have been already published
(Pichardo \et 2003; Martos \et 2004; Antoja \et 2009; Antoja \et
2011).

\begin{figure}
\includegraphics[width=1\textwidth]{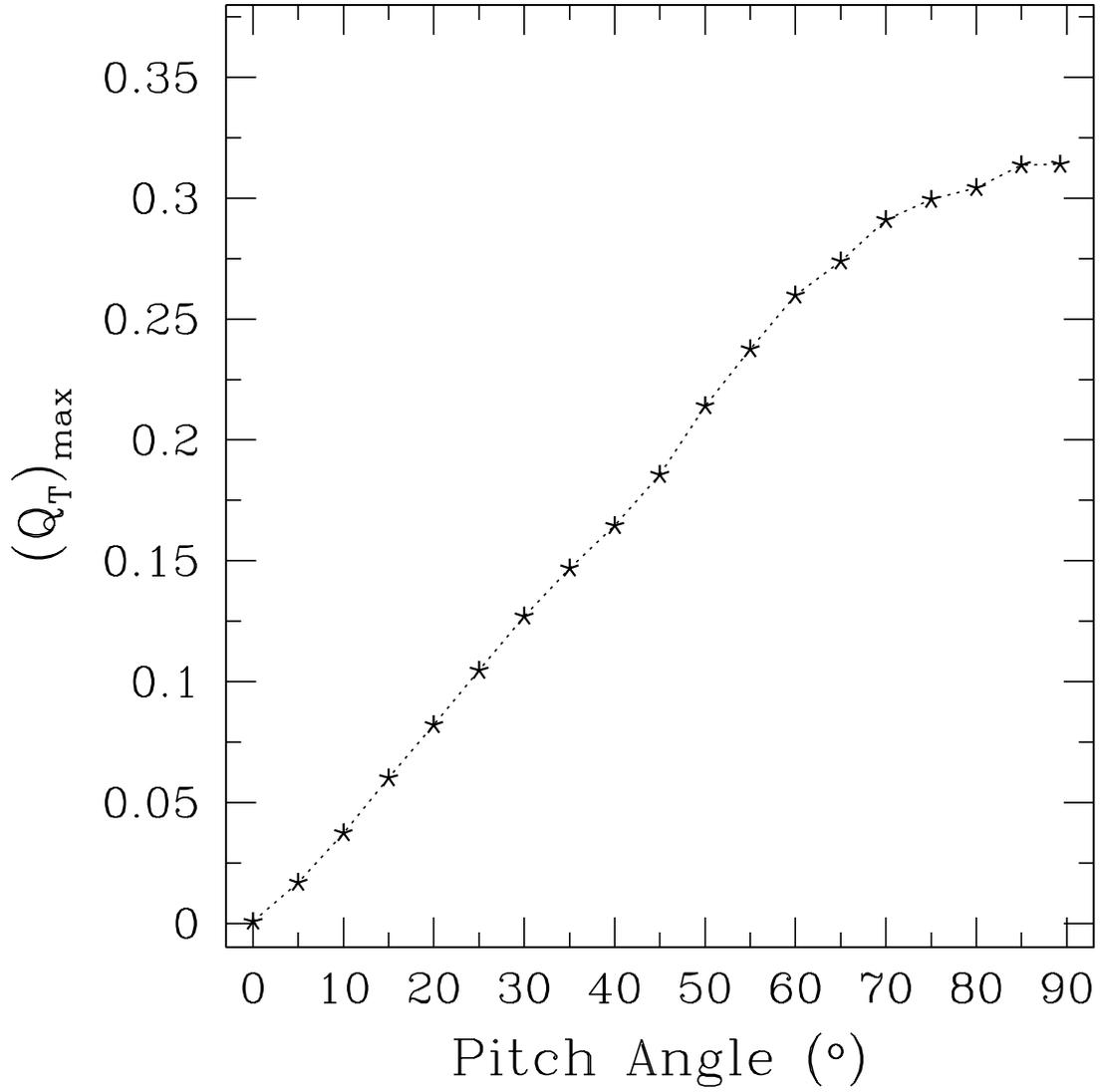}
\caption{(Q$_{\rm T})_{\rm max}$ represents the maximum value of the
  parameter Q$_T(R)$ (maximum relative torques) vs. pitch angle of the
  spiral arms model.}
\label{parameterQ}
\end{figure}

The corresponding parameters used to produce models for late type
spiral galaxies are presented in Table \ref{tab:parameters}.  Spiral
arms self-consistency has been tested through the reinforcement of the
spiral potential by stellar orbits (Patsis \et 1991; Pichardo \et
2003). The total mass of the spiral arms in our model is of 3\% of the
disc mass, which represents conservative (low mass) spiral arms for
late type disc galaxies. We have employed the known parameter Q$_T(R)$
(Combes \& Sanders 1981) applied to bars and spiral arms (Buta \&
Block 2001; Laurikainen \& Salo 2002; Vorobyov 2006; Kalapotharakos et
al. 2010, etc.), to measure the strength of the spiral arms ({\tt
  PERLAS}) in order to compare with observations and other models. We
present in Figure \ref{parameterQ} the maximum value, for models with
different pitch angles, of the parameter Q$_{\rm T}$(R)=F$_{\rm
  T}^{\rm max}$(R)/$|\langle$F$_{\rm R}$(R)$\rangle|$, where F$^{\rm
  max}_{\rm T}$ =$|\left(\frac{1}{R} \ \partial\Phi ({\rm
  R},\theta)/\partial \theta\right)|_{\rm max}$, represents the
maximum amplitude of the tangential force at radius R, and
$\langle$F$_{\rm R}$(R)$\rangle$, is the mean axisymmetric radial
force at the same radius, derived from the m=0 component of the
gravitational potential. Each point in this curve represents a
different pitch angle, going from 0$^\deg$ to almost
90$^\deg$. Maximum values up to 0.4 for this parameter are reasonable
for late spirals (Buta \et 2005).

The axisymmetric potential includes a Miyamoto-Nagai disk and bulge,
and a massive halo (based on the potential described by Allen \&
Santill\'an 1991 for the Milky Way) to fit known observational
parameters on late spirals: mainly rotation curves, mass ratios
(between the components: bulge, disk and halo) and scale lengths. In
Table \ref{tab:parameters}, we present the employed axisymmetric and
non-axisymmetric (spiral arms) parameters.

\begin{deluxetable}{lll}
\tablecolumns{2}
\tablewidth{0pt}
\tablecaption{Parameters of the Spiral Arms Models}
\tablehead{{Parameter} &{Value}& {Reference}}
\startdata

 &{\it Spiral Arms}  \\
\hline
locus             & Logarithmic  & 1,10\\
arms number       & 2  & 2\\
pitch angle       & 10-60$\deg$  & 3,8 \\
M$_{Spiral}$/M$_{disk}$& 3\%  \\
scale-length      & disk based: 3 $\kpc$   & 4,5\\
Radial force contrast    & 5-10\% & 6 \\
pattern speed ($\Omega_{sp}$) & -20 (clockwise) $\kmskpc$   & 1,7 \\
ILR position    & 2.03 kpc & \\
Corotation position    & 8.63 kpc & \\
inner limit       & 2.03 kpc & $\sim$ILR position based\\
external limit    & 8.63 kpc & $\sim$corotation position based\\
\hline
 &{\it Axisymmetric Components}  \\
\hline
Disk Mass / Halo Mass & 0.1 (up to 100 kpc halo radius) &  4,9  \\
Bulge Mass / Disc Mass & 0.2 & 5,9 \\
Rot. Curve (V$_{max}$)&  170 $\kms$& 8  \\
Disk Mass         & $5.10 \times^{10}$ M$_\odot$& 4  \\
Bulge Mass        & $1.02 \times^{10}$ M$_\odot$& $M_D/M_B$ based  \\
Halo Mass         & $4.85 \times^{11}$ M$_\odot$& $M_D/M_H$ based  \\
Disk scale-length & 3 $\kpc$   & 4,5
\label{tab:parameters}
\enddata
\tablerefs{ 1)~Grosbol \& Patsis 1998.
            2)~Drimmel \et 2000.
            3)~Kennicutt 1981.
            4)~Pizagno \et 2005
            5)~Weinzirl \et 2009.
            6)~Contopoulos 2007.
            7)~Patsis \et 1991; Fathi \et 2009.
	    8)~Ma \et 2000; Brosche 1971; Sofue \& Rubin 2001.
            9)~Block \et 2002.
           10)~Pichardo \et 2003.
}
\end{deluxetable} 


\section{Results}\label{results}

\begin{figure}
\includegraphics[width=1\textwidth]{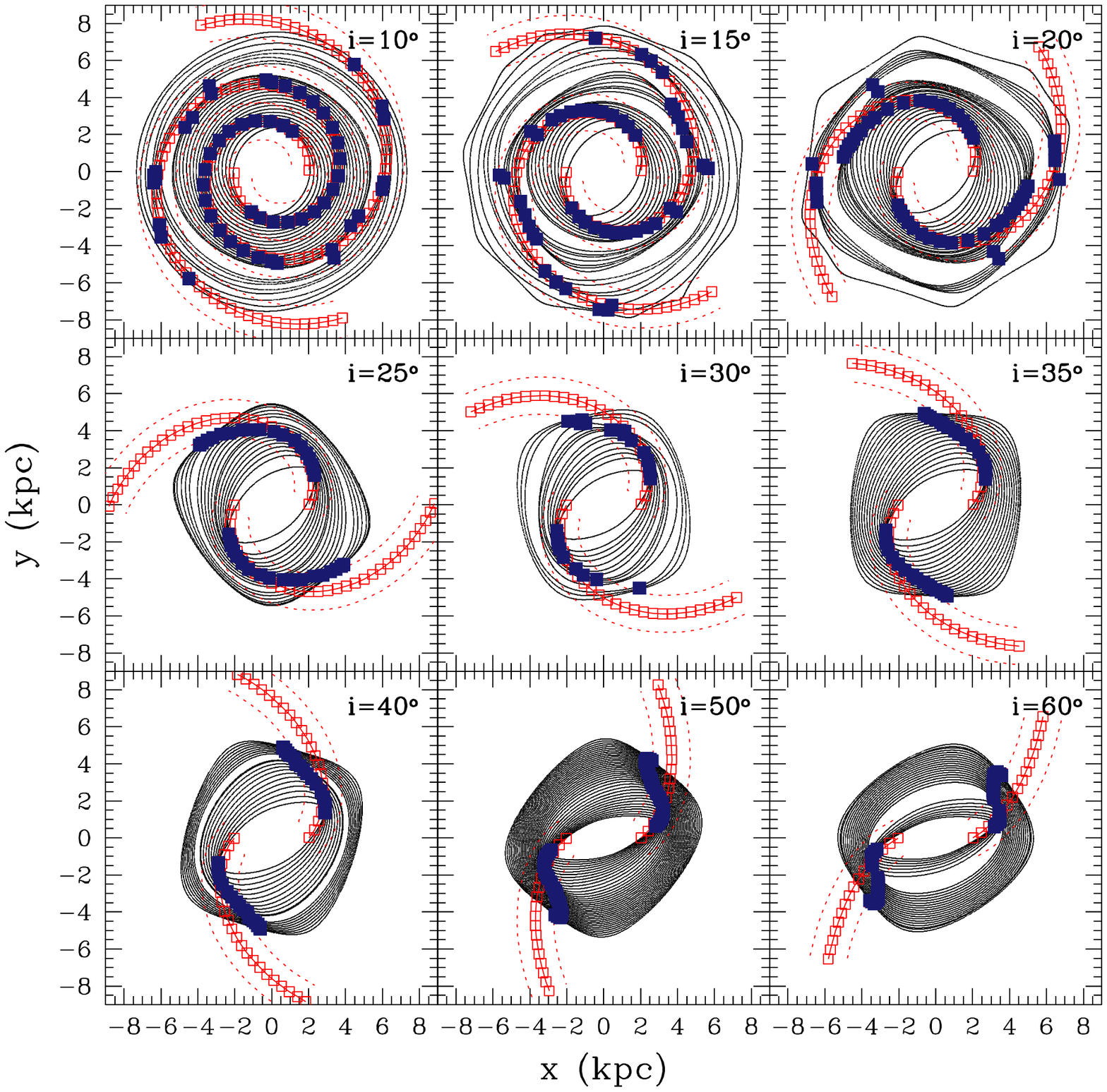}
\caption{Periodic orbits and response maxima (filled squares) in
  models with the spiral loci (open squares) for the 3D spiral model
  (Table \ref{tab:parameters}), with pitch angles ranging from 10 to
  60 $\deg$.}
\label{periodicas}
\end{figure}

In the case of long lasting steady potentials, self-consistency can be
tested undirectly through the construction of periodic orbits. The
existence of periodic orbits supporting a large scale structure such
as spiral arms increases the probability of mantaining long lasting
large scale structures. We present a periodic orbit study in Figure
\ref{periodicas}. In order to quantify the support of periodic orbits
to the spiral arm potential, we follow the method of Contopoulos \&
Grosb\o l (1986) to obtain the density response to the given spiral
perturbation. This method assumes that the stars with orbits trapped
around an unperturbed circular orbit, and with the sense of rotation
of the spiral perturbation, are also trapped around the corresponding
central periodic orbit in the presence of the perturbation. In this
manner, we computed a series of central periodic orbits and found the
density response along their extension, using the conservation of mass
flux between any two successive orbits. We found the position of the
density response maxima (filled squares in Figure \ref{periodicas})
along each periodic orbit, thus the positions of the response maxima
on the galactic plane are known. These positions are compared with the
center of the imposed spiral arms potential, i.e. the spiral locus
(open squares in Figure \ref{periodicas}).

For smaller pitch angles ($i \la 20\deg$), the density response maxima
support closely the spiral arms potential, making the existence of
long lasting spiral arms more probable. On the other hand, for spiral
arms with pitch angles larger than $\sim 20\deg$, the response maxima
precede systematically the spiral arms potential, i.e. the response
produces spiral arms with much smaller pitch angles than the spiral
arms potential. The response (or support) ``avoids'' open long lasting
spiral arms. Beyond 20$\deg$ of pitch angle for the spiral potential,
the density response keeps almost the same pitch angle (approximately
18$\deg$ to 22$\deg$) independently of the potential. Long lasting
spirals are not supported anymore, spiral arms in this case may be
rather transient.

\begin{figure}
            \includegraphics[width=1\textwidth]{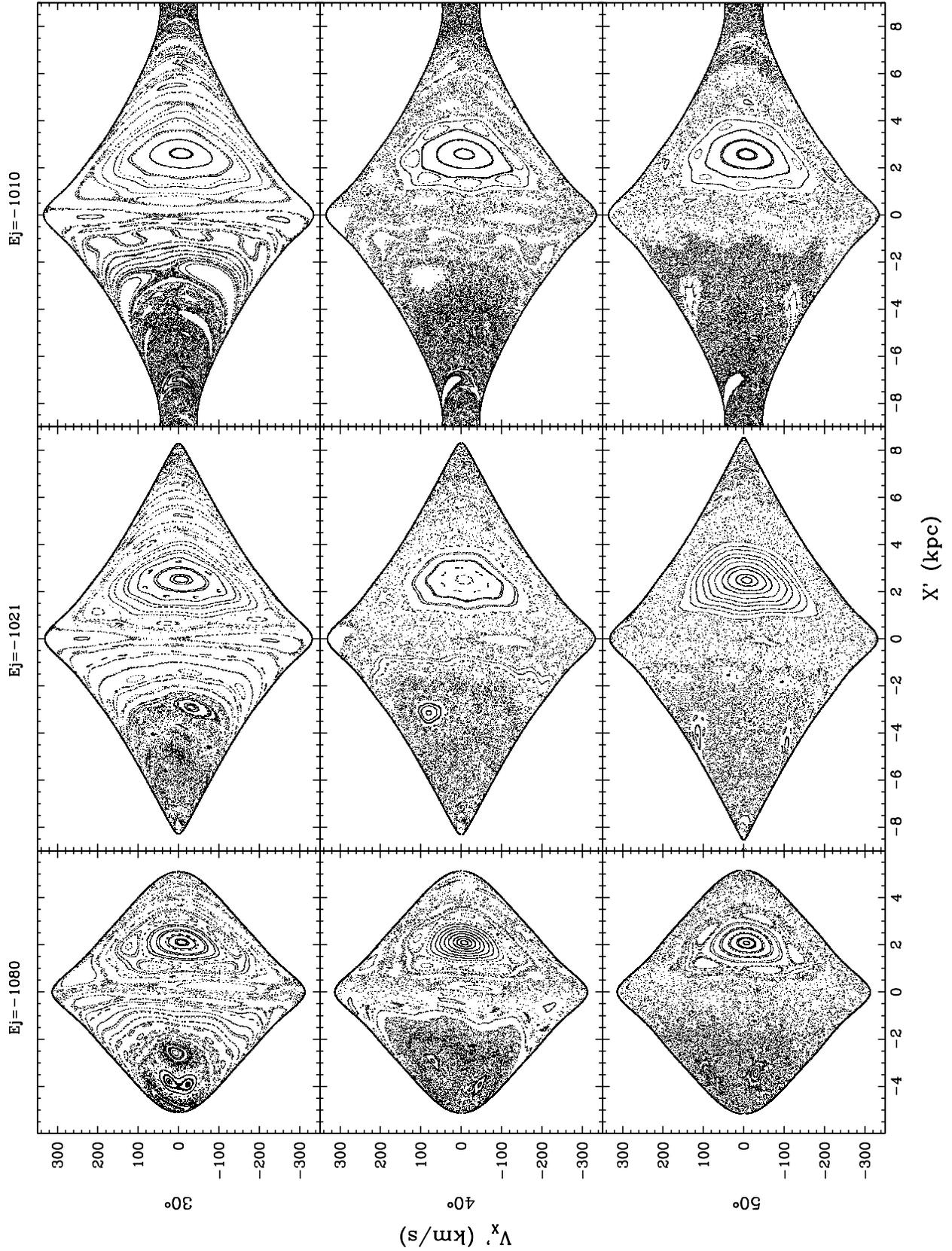}
\caption{Phase space diagrams with E$_J= [-1080,-1010]$, in units of
  $10^2$ km$^2$ s$^{-2}$. From upper to bottom lines of diagrams,
  pitch angles go from 30$\deg$ to 50$\deg$.}
\label{arms}
\end{figure}

In Figure \ref{arms} we present a 3$\times$3 phase space diagrams
mosaic to show the main results. These 9 panels show Poincar\'e
diagrams with different Jacobi energy families running from Ej=-1080
to -1010 $\times 10^2$ km$^2$ s$^{-2}$, covering the total extension
of the spiral arms, as we go from 30$\deg$ (upper line of diagrams) to
50$\deg$ (bottom line of diagrams). The left part of each diagram
represents prograde orbits in the reference system of the spiral arms.

Observations of spiral galaxies show pitch angles up to $\sim 50\deg$
(Ma \et 2000). Despite that periodic orbits are not supporting spiral
arms, the existence of very open spirals, could indicate a probable
transient nature. However, even then, some ordered orbits are expected
to support for short periods these large scale structures.

We have studied the effect of increasing the pitch angle in a typical
late type spiral galaxy model. With this study we search for a limit
for the pitch angle, for which chaos becomes pervasive and destroys
all ordered orbits in the relevant spiral arms region. At a pitch
angle of 20$\deg$ or less, the majority of orbits are ordered and
simple, periodic orbits support spiral arms up to close to
corotation. As we increase the pitch angle, at 30$\deg$ (first line of
diagrams from the top of Figure \ref{arms}), the orbital behavior is
much more complex, presenting resonant islands and the onset of chaos
is clear, surrounding the stable periodic orbits, yet supporting them
in a contained region. For 40$\deg$ (second line of diagrams), chaos
becomes pervasive compromising the available phase space around the
stable periodic orbits. For 50$\deg$ (bottom line of diagrams), the
chaotic region covers almost all regular prograde orbits, approaching
closely to the main periodic orbits. For pitch angles beyond $\sim
50\deg$ chaos destroys periodic orbits.

In all cases reported in this work, the radial position of corotation
was kept fixed (see Table \ref{tab:parameters}). However, taking into
account earlier work by Contopoulos \& Grosb\o l (1986); Patsis,
Contopoulos \& Grosb\o l (1991), among others, we produced several
experiments taking the position of corotation (to 16.5 kpc by reducing
the spiral arms angular speed from 20 to 10 km s$^{-1}$ kpc$^{-1}$)
considerably far from the end of the spiral arms to check whether it
is relevant in the observed chaotic behavior. Although a large
fraction of chaos is indeed produced close to the corotation
resonance, we found that even with corotation far from the end of
spiral arms, a fraction of chaos is generated toward larger pitch
angles, although not enough to compromise ordered regions around the
periodic orbits. We then produced experiments where we fixed the end
of the spiral arms close to the corotation position, and changed only
one pararameter: the pitch angle. We found then that the chaos
produced in corotation becomes much stronger as we increase the pitch
angle. Chaos seems a combination between the effect of the corotation
resonance and the pitch angle.

\section{Conclusions}\label{conclusions}

With the use of an axisymmetric fixed model to simulate a typical late
type galaxy as a background, we superposed a bisymmetric steady spiral
arm potential ({\tt PERLAS}) and studied the evolution of orbital
behavior in the plane of the disk, as we change the pitch angle going
from 10$\deg$ to 60$\deg$, in order to set some structural
restrictions to the spiral arms, based on orbital dynamics. Observed
galaxies classified as late type spirals present a wide scatter in
pitch angles, going from $\sim 10\deg$ to $50\deg$.  With these family
of models, we have carried out an exhaustive orbital study (order and
chaos) with periodic orbits and with phase space diagrams.

In this paper we present the first restriction relative to the pitch
angle. In the case of ordered motion, with periodic orbits, a limit in
the pitch angle of the density response is found at approximately
$20\deg$, up to which, the density response reinforces the spiral arms
potential at all radii, i.e. with a more long-lasting nature. Beyond
this limit, the density response ``avoids'' to follow the spiral arm
potential, producing pitch angles much smaller than the background
spiral potential. Spiral arms beyond this limit might be better
explained as transient structures.

A second restriction is obtained out of chaos behavior. With the phase
space orbital study, going from pitch angles of 10$\deg$, where order
reigns, to more than 50$\deg$, where chaos becomes pervasive. With
this orbital study we are able to set a limit value for the maximum
pitch angle before the system becomes completely chaotic. This limit
closely coincides with the observational maximum pitch angle of
spirals ($\sim 50\deg$), suggesting a possible relation between the
structural characteristics of the galaxy and chaos.

\acknowledgments It is a pleasure to acknowledge Panos Patsis and the
anonymous referee for enlightening comments that considerably improved
this work. We thank PAPIIT through grants IN110711-2, IN119306 and
IN-109509.

\end{document}